\documentclass[aps,floatfix,showpacs,preprintnumbers,amsmath,amssymb,a4paper,preprint]{revtex4}
\usepackage[dvips]{graphicx}
\usepackage{bm}

\begin{document}

\title{Hamiltonian description of a self-consistent interaction between charged particles and electromagnetic waves }
\author{R. Bachelard}
\author{C. Chandre}
\author{M. Vittot}

\affiliation{Centre de Physique Th\'eorique\footnote{UMR 6207 of the CNRS, Aix-Marseille and Sud Toulon-Var Universities. Affiliated with the
CNRS Research Federation FRUMAM (FR 2291). CEA registered research laboratory LRC DSM-06-35.} - CNRS, Luminy - Case 907, 13288 Marseille
cedex 09, France}

\date{\today}

\begin{abstract}
The Hamiltonian description of the self-consistent interaction between an electromagnetic plane-wave and a co-propagating beam of charged particles is considered. We show how the motion can be reduced to a one-dimensional Hamiltonian model (in a canonical setting) from the Vlasov-Maxwell Poisson brackets. The reduction to this paradigmatic Hamiltonian model is performed using a Lie algebraic formalism which allows us to remain Hamiltonian at each step of the derivation.
\end{abstract}


\maketitle

\section{Introduction}

The interaction between electromagnetic fields and a beam of charged particles exhibit a dynamics which is nowadays not fully understood, mainly due to the complexity inherent to the infinite dimensional phase space. Although some attention has been devoted to controlling those systems, a preliminary step is to shed light on the dynamics by analysing for instance phase space structures and transport properties. In this ambitious framework, reduced models have proved to be very valuable for this task. 
The reduction is of course guided by the physics of a particular setting. For instance, a beam of ultra-relativistic electrons interacts with plane waves (whether these are external ones or produced by the beam moving into an array of magnets, called undulator) has shown to be a way for producing a coherent light source. In what follows, we consider such a beam of electrons moving inside an undulator which produces a static (but non-uniform) magnetic field. The acceleration produced by the external magnetic field makes them emit a synchrotron radiation, which is self-consistently interacting with the particles. Under some resonance condition, the intensity of this electromagnetic wave grows exponentially and then saturates. In order to capture this effect, a one-dimensional Hamiltonian model has been proposed. The reduced Hamiltonian describes the evolution of the position $\theta_j$ and relative momentum $p_j$ (around a resonant value) of the $j$-th particle. The $N$ particles interact through a wave described by its intensity $I$ and phase $\phi$. It reads~:
\begin{equation}
\label{eqn:bonif}
H=\sum_{j=1}^{N}\left(\frac{p_j^2}{2}+2\sqrt{I}\sin(\theta_j-\phi)\right),
\end{equation}
where $(\theta_j,p_j)$ and $(\phi,I)$ are canonical pairs of conjugate variables. 

In the Free Electron Laser (FEL) configuration, this model has been derived \cite{bonifacio,bonham} from the equations of motion of charged particles and Maxwell's equations. Some approximations were involved during the course of the derivation, and were guided by the physics of the device, and among them, a specific form for the radiated fields (as a plane wave), a one-dimensional reduction (obtained by reducing the dynamics in the transverse plane), an expansion around a resonance fixed by the characteristics of the undulator. In addition, this Hamiltonian model was also proposed to describe the wave-particle interaction in other contexts, such as the beam-plasma instability \cite{mynick,escande}, or the Collective Atomic Recoil Laser \cite{carl}.

In this article, we propose a derivation of the one-dimensional Hamiltonian~(\ref{eqn:bonif}) for the self-consistent interaction from the Vlasov-Maxwell equations in a Hamiltonian setting. This allows us to show that at each step of the derivation, the Hamiltonian structure of the problem is conserved. Another advantage is that the conserved quantity (namely the total momentum of the system) is also easily deduced from the conserved quantity of the Vlasov-Maxwell equations. Our approach follows from this algebraic framework: Instead of working with the equations of motion for the derivation, we consider the Hamiltonian and its associated Poisson bracket (for an introduction, see Refs.~\cite{cary,dragt}). By using a canonical version, the Poisson bracket remains canonical (or generalized canonical in a broader sense). Therefore the main approximations and computations have to be done on a scalar function, the Hamiltonian which aims at simplifying the derivation.   

In Sec.~\ref{sec:bla} we recall some basics of the Hamiltonian formulation of the Vlasov-Maxwell equations for a continuous description of the particle distribution. In Sec.~\ref{sec:deriv}, we apply to this Hamiltonian system the approximations and reduction necessary for the derivation of the reduced model~(\ref{eqn:bonif}) by expressing first the Vlasov-Maxwell system into a canonical setting (Sec.~\ref{sec:can}), then performing the one-dimensional reduction (Sec.~\ref{sec:one}), changing the reference frame (Sec.~\ref{sec:frame}), and expanding the resulting Hamiltonian around a resonance condition (Sec.~\ref{sec:reso}). Finally,  we deduce (Sec.~\ref{sec:cons}) a conserved quantity of the reduced model~(\ref{eqn:bonif}) from a conserved quantity of the Vlasov-Maxwell equations by following the same procedure as for the reduction of the Hamiltonian.

\section{Hamiltonian formulation of Vlasov-Maxwell equations} 
\label{sec:bla}

The dynamics of Hamiltonian~(\ref{eqn:bonif}) follows from Hamilton's equations for each pair of canonically conjugate variables. More generally the dynamics of an observable $F$ (function of phase space coordinates $(\{\theta_i,p_i\},I,\phi)$) is given by~:
\begin{equation}\label{evol} \frac{dF}{dt}=\{ H,F \}, \end{equation}
where the Poisson bracket between two observables is given by
$$
\{F,G\}=\sum_{i=1}^N\left( \frac{\partial F}{\partial p_i}\frac{\partial G}{\partial \theta_i}-\frac{\partial F}{\partial \theta_i}\frac{\partial G}{\partial p_i}\right)+\frac{\partial F}{\partial I}\frac{\partial G}{\partial \phi}-\frac{\partial F}{\partial \phi}\frac{\partial G}{\partial I}.
$$ 
In a continuous setting, this Hamiltonian model can be extended in a straightforward way. The beam is now described by a distribution function $f(\theta,p)$ which constitutes a dynamical field, i.e., for each location in phase space $(\theta,p)$, the density of particles $f(\theta,p)$, labelled by the phase space coordinates of the particles, evolves dynamically. The one-dimensional Hamiltonian model is generalized from Hamiltonian~(\ref{eqn:bonif})
\begin{equation}
\label{eqn:boni}
H[f,I,\phi]=\iint d\theta dp f(\theta,p) \left[\frac{p^2}{2}+2\sqrt{I}\sin{(\theta-\phi)}\right],
\end{equation}
where the dynamical variables are now $I$ and $\varphi$, and a field of variables $f(\theta,p)$. The dynamics of $f$, $I$ and $\theta$ are obtained using the canonical Poisson bracket:
\begin{equation} \{ F,G \} = \iint d\theta dp f(\theta,p) \left[\frac{\partial}{\partial p} \frac{\delta F}{\delta f} \frac{\partial}{\partial \theta} \frac{\delta G}{\delta f} - \frac{\partial}{\partial \theta} \frac{\delta F}{\delta f} \frac{\partial}{\partial p} \frac{\delta G}{\delta f}\right] + \frac{\partial F}{\partial I} \frac{\partial G}{\partial \phi}-\frac{\partial F}{\partial \phi} \frac{\partial G}{\partial I}, 
\label{eqn:braboni}
\end{equation}
i.e.\ it leads to a Vlasov equation for $f$
$$
\frac{\partial f}{\partial t}+p\frac{\partial f}{\partial \theta}-2\sqrt{I}\cos(\theta-\phi)\frac{\partial f}{\partial p}=0,
$$
where we notice that $df/dt=\partial f/\partial t$ since we use an Eulerian description for the observables. This equation has been used to estimate quantitatively some features of the dynamics, like the derivation of a reduced dimensional model~\cite{tennyson} or the characteristics of the bunching in the saturated regime~\cite{anto05,anto06}.

If $f$ is a Klimontovitch distribution, that is the distribution function is a sum of Dirac representing some point particles:
$$ f(\theta,p)=\sum_j \delta(\theta-\theta_j(t))\delta(p-p_{j}(t)), $$ 
we recover the equations for Hamiltonian~(\ref{eqn:bonif}). 

The continuous formalism~(\ref{eqn:boni})-(\ref{eqn:braboni}) is particularly well-suited for an algebraic treatment of the dynamics (see e.g. \cite{redmars}). In what follows we use Vlasov-Maxwell equations to derive Hamiltonian~(\ref{eqn:boni}). In order to do this we use a Hamiltonian formulation of these equations. First let us recall that the interaction between electromagnetic fields and charged particles (of normalized mass $m=1$ and charge $e=1$) is given  as the sum of the kinetic energy of the particles plus the energy of the field~\cite{Morrison,marsden,bialynicki}:
\begin{equation}\label{ham0} H = \iint d^3q d^3p f({\bf q},{\bf p})\sqrt{1 + {\bf p}^2}+ \int d^3q \frac{|{\bf E}({\bf q})|^2 + |{\bf B}({\bf q})|^2}{2}, \end{equation}
where $f({\bf q},{\bf p})$ describes the distribution of particles in phase-space. Even though the kinetic energy of the particles and the electromagnetic energy appear to be decoupled in the Hamiltonian, the interaction between the matter and the fields comes from the bracket which gives the dynamics~:
\begin{eqnarray}
 \{ F,G \} &=& \iint d^3 q d^3 p \, f \left[\frac{\partial}{\partial{\bf p}} \frac{\delta F}{\delta f} \cdot \frac{\partial}{\partial{\bf q}} \frac{\delta G}{\delta f} - \frac{\partial}{\partial{\bf q}} \frac{\delta F}{\delta f} \cdot \frac{\partial}{\partial{\bf p}} \frac{\delta G}{\delta f}\right] \nonumber \\
&&- \iint d^3 q d^3 p \, f {\bf B}\cdot\left[ \frac{\partial}{\partial{\bf p}} \frac{\delta F}{\delta f} \times \frac{\partial}{\partial{\bf p}} \frac{\delta G}{\delta f}\right]\nonumber \\
&&+ \iint d^3 q d^3 p \left[ \frac{\delta F}{\delta f} \frac{\partial f}{\partial {\bf p}} \cdot \frac{\delta G}{\delta {\bf E}} -\frac{\delta G}{\delta f} \frac{\partial f}{\partial {\bf p}} \cdot \frac{\delta F}{\delta {\bf E}}\right]\nonumber \\
&&+ \int d^3 q \left[\left(\nabla \times \frac{\delta F}{\delta {\bf B}}\right) \cdot \frac{\delta G}{\delta {\bf E}} -\frac{\delta F}{\delta {\bf E}} \cdot \left(\nabla \times \frac{\delta G}{\delta {\bf B}}\right)\right].\label{brac0} \end{eqnarray}
This bracket satisfies the antisymmetry property, the Leibnitz product rule and the Jacobi identity. Here, the Lie algebra on which this bracket operates is the set of smooth functionals $F[f({\bf q},{\bf p}),{\bf E}(\bf q),{\bf B}(\bf q)]$. Using Hamiltonian~(\ref{ham0}) and the brackets~(\ref{brac0}), Eq.~(\ref{evol}) allows one to retrieve Maxwell's equations for ${\bf E}$ and ${\bf B}$, as well as Vlasov equation for $f$~:
\begin{eqnarray}
&& \frac{\partial f}{\partial t}\equiv\dot{f}=\{H,f\}=-{\bf v}\cdot\nabla f-({\bf E+{\bf v}\times{\bf B}})\cdot \frac{\partial f}{\partial {\bf p}},\nonumber\\
&& \frac{\partial {\bf E}}{\partial t}\equiv\dot{\bf E}=\{H,{\bf E}\}= \nabla \times {\bf B}-\int d^3p {\bf v} f,\nonumber\\
&& \frac{\partial {\bf B}}{\partial t}\equiv\dot{\bf B}=\{H,{\bf B}\}= -\nabla \times {\bf E},\nonumber
\end{eqnarray} 
where ${\bf v}$ is the velocity
\begin{equation} {\bf v}=\frac{\bf p}{\sqrt{1+{\bf p}^2}}. \label{eq:velocity}\end{equation}
We notice that the first line of the right hand side of Eq.~(\ref{brac0}) refers only to the particles (and it is canonical), the second and third lines are the field-particle interaction terms (non-canonical terms) and the last line is a field-only term (which is also canonical).

\section{Interaction between a plane wave and a co-propagating beam of particles}
\label{sec:deriv}

\subsection{Expression of the Hamiltonian system in a canonical way}
\label{sec:can}

The Vlasov-Maxwell equations~(\ref{ham0})-(\ref{brac0}) can also be described using the potentials instead of the fields~\cite{marsden}. The Lie algebra is now a set of functionals $F[f_{\rm mom}({\bf q},{\bf p}),{\bf A}({\bf q}),{\bf Y}({\bf q})]$. The Hamiltonian and the bracket become~:
\begin{eqnarray}\label{ham1} &&H = \iint d^3q d^3p f_{\rm mom}\sqrt{1 + ({\bf p}-{\bf A})^2}+ \int d^3q \frac{|{\bf Y}|^2 + |\nabla \times {\bf A}|^2}{2}, \\
 &&\{ F,G \} = \iint d^3 q d^3 p \, f_{\rm mom} \left[\frac{\partial}{\partial{\bf p}} \frac{\delta F}{\delta f_{\rm mom}} \cdot \frac{\partial}{\partial{\bf q}} \frac{\delta G}{\delta f_{\rm mom}} - \frac{\partial}{\partial{\bf q}} \frac{\delta F}{\delta f_{\rm mom}} \cdot \frac{\partial}{\partial{\bf p}} \frac{\delta G}{\delta f_{\rm mom}}\right] \nonumber \\
&&\quad \qquad + \int d^3q \left( \frac{\delta F}{\delta{\bf Y}}\cdot  \frac{\delta G}{\delta {\bf A}}-\frac{\delta F}{\delta{\bf A}}\cdot  \frac{\delta G}{\delta {\bf Y}}\right).\label{brac1}\end{eqnarray}
This can be obtained from Eqs.~(\ref{ham0})-(\ref{brac0}) using the change of coordinates
\begin{eqnarray*}
&& f({\bf q},{\bf p})=f_{\rm mom}({\bf q},{\bf p}+{\bf A}),\\
&& {\bf E}=-{\bf Y},\\
&& {\bf B}=\nabla \times {\bf A}.
\end{eqnarray*}
We notice that this time, the Poisson bracket is canonical and there is no term in this bracket which couples the particles and the field. However, the coupling term is present in Hamiltonian~(\ref{ham1}). \\

We translate the potential vector by a static ${\bf A}_w({\bf q})$, which is imposed externally (as in an undulator). We notice that a translation of ${\bf A}$ by a quantity ${\bf A}_w$ is a canonical transformation, which implies that the bracket~(\ref{brac1}) is not changed. The new Hamiltonian reads~:
$$ H = \iint d^3q d^3p f_{\rm mom}\sqrt{1 + ({\bf p}-{\bf A}_w-{\bf A})^2}+ \int d^3q \frac{|{\bf Y}|^2 +2 \nabla\times {\bf A}_w\cdot \nabla \times {\bf A}+ |\nabla \times {\bf A}|^2}{2}, $$
where we have dropped the constant quantity $\int d^3q |\nabla \times {\bf A}_w|^2/2$. In particular, we notice that the dynamics of the radiated field is
$$
\dot{\bf A}=\{H,{\bf A}\}=\frac{\delta H}{\delta \bf Y}={\bf Y},
$$
which is equivalent to the equation for the radiated electric field ${\bf E}_r$:
$$
{\bf E}_r=-\frac{\partial {\bf A}}{\partial t}.
$$
For a wave co-propagating with the electrons in the $z$-direction, one can define the $k$-mode of the wave as follows:
$$ {\bf A}_k({\bf q_\perp})=\frac{1}{L}\int dz e^{-ikz} {\bf A}({\bf q}), $$
$$ {\bf Y}_k({\bf q_\perp})=\frac{1}{L}\int dz e^{-ikz} {\bf Y}({\bf q}), $$
where ${\bf q_\perp}=(x,y)$ and $L$ is the length of the cavity where the interaction takes place. This gives the Fourier expansion in the propagation direction ${\bf Y}({\bf q})=\sum_{k} {\bf Y}_k({\bf q_\perp}){\rm e}^{i k z}$ and ${\bf A}({\bf q})=\sum_{k} {\bf A}_k({\bf q_\perp}){\rm e}^{i k z}$. Furthermore, since
$$ \frac{\delta {\bf A}_k({\bf q_\perp})}{\delta {\bf A}({\bf q}')}=\frac{1}{L}e^{-ikz}\delta({\bf q_\perp}-{\bf q_\perp'}) ,$$
which is obtained from the definition and linearity of the functional derivative,
it follows from the bracket~(\ref{brac1}) that
$$ \{ {\bf Y}_k({\bf q_\perp}),{\bf A}_{-k'}({\bf q_\perp'}) \} =\frac{1}{L} \delta_{k k'} \delta({\bf q_\perp}-{\bf q_\perp'}). $$
Since we also have:
$$ \{ {\bf Y}_k({\bf q_\perp}),{\bf Y}_{k'}({\bf q_\perp'}) \} = \{ {\bf A}_k({\bf q_\perp}),{\bf A}_{k'}({\bf q_\perp'}) \} =0,$$
the field part of the bracket turns into  
$$
\int d^3q \left( \frac{\delta F}{\delta{\bf Y}}\cdot  \frac{\delta G}{\delta {\bf A}}-\frac{\delta F}{\delta{\bf A}}\cdot  \frac{\delta G}{\delta {\bf Y}}\right)=\frac{1}{L}\sum_{k}\int d^2q_\perp \left( \frac{\delta F}{\delta{\bf Y}_k}\cdot  \frac{\delta G}{\delta {\bf A}_{- k}}-\frac{\delta F}{\delta{\bf A}_k}\cdot  \frac{\delta G}{\delta {\bf Y}_{- k}}\right).
$$
We now consider the {\it paraxial approximation} both for the radiated and external fields, i.e. we neglect their spatial variations in the $x$ and $y$ directions, so that they are homogeneous in the transverse section $S$ of interaction, and null outside it. We notice that this paraxial approximation is the strongest approximation involved in the derivation process. The dimensional reduction crucially depends on it. In addition to this approximation, we restrict the derivation to a {\it monochromatic} wave, i.e. we only take into account one Fourier mode $k$ in the propagation direction, and consider the case of a {\it circularly polarized} radiated wave. Other modes can be included in the derivation in a very similar way, but we have only kept one mode for the sake of clarity of the derivation. These two approximations allow us to define the complex amplitude of the wave $a$ such that
\begin{eqnarray} 
&& {\bf A}=-\frac{i}{\sqrt{2}}\left[ ae^{ikz}{\bf \hat{\mathrm e}}-a^* e^{-ikz}{\bf \hat{\mathrm e}^*} \right],\nonumber \\
&& {\bf Y}=-\frac{k}{\sqrt{2}}\left[ ae^{ikz}{\bf \hat{\mathrm e}}+a^* e^{-ikz}{\bf \hat{\mathrm e}^*} \right], \label{approxwave}
\end{eqnarray}
with ${\bf \hat{\mathrm e}}=({\bf \hat{x}}+i{\bf \hat{y}})/\sqrt{2}$. Conversely, $a$ can be defined as
$$ a=\frac{1}{kV}\int d^3q e^{-ikz} (-{\bf Y}+ik{\bf A})\cdot{\bf \hat{\mathrm e}^*}=\frac{1}{kS}\int d^2{\bf q_\perp} (-{\bf Y}_k+ik{\bf A}_k)\cdot{\bf \hat{\mathrm e}^*} , $$
where $V$ is the volume of the interaction domain, i.e. $V=LS$ with the above notations. Then, since
\begin{eqnarray*} 
&& \frac{\delta a}{\delta {\bf Y}_k ({\bf q_\perp})}=-\frac{1}{kS}\ {\bf \hat{\mathrm e}^*}, \\
&& \frac{\delta a}{\delta {\bf A}_k ({\bf q_\perp})}=\frac{i}{S}\ {\bf \hat{\mathrm e}^*},
\end{eqnarray*}
we get $$\{ a,a^* \} =\frac{i}{kV},$$ so that $a$ and $a^*$ are the new conjugate variables describing the radiated field. Hence, the bracket turns into:
\begin{eqnarray*} 
\{ F,G \} &=& \iint d^3q d^3p f_{\rm mom} \left[\frac{\partial}{\partial{\bf p}} \frac{\delta F}{\delta f_{\rm mom}} \cdot \frac{\partial}{\partial{\bf q}} \frac{\delta G}{\delta f_{\rm mom}} - \frac{\partial}{\partial{\bf q}} \frac{\delta F}{\delta f_{\rm mom}} \cdot \frac{\partial}{\partial{\bf p}} \frac{\delta G}{\delta f_{\rm mom}}\right] \\
&& + \frac{i}{kV} \left(\frac{\partial F}{\partial a} \frac{\partial G}{\partial a^*}-\frac{\partial F}{\partial a^*} \frac{\partial G}{\partial a}  \right). \end{eqnarray*}
As for the Hamiltonian, since we have $\nabla \times {\bf A}=k{\bf A}$ and $|{\bf Y}|^2=k^2 |{\bf A}|^2=k^2 aa^*$ for the vector potential ${\bf A}$, the energy of the radiated wave now reads:
$$  \int d^3q \frac{|{\bf Y}|^2 + |\nabla \times {\bf A}|^2}{2}=k^2 V a a^*, $$
where we have used the relations ${\bf \hat{\mathrm e}}\cdot{\bf \hat{\mathrm e}}=0$ and ${\bf \hat{\mathrm e}}\cdot{\bf \hat{\mathrm e}^*}=1$.
So that the Hamiltonian becomes:
\begin{eqnarray*} H=&& \iint d^3q d^3p f_{\rm mom}\Bigg[ 1 + {\bf p}^2 +a a^* -i\sqrt{2}(a e^{ikz}{\bf \hat{\mathrm e}}-a^* e^{-ikz}{\bf \hat{\mathrm e}^*}) \cdot {\bf A}_w
\\ &&+\vert {\bf A}_w\vert^2-2 {\bf p}_\perp \cdot ({\bf A}_w+{\bf A})\Bigg]^{1/2} \\
 &&+ k^2 V a^* a -\frac{i kS}{\sqrt{2}} \int dz (a e^{ikz}{\bf \hat{\mathrm e}}-a^* e^{-ikz}{\bf \hat{\mathrm e}^*})\cdot (\nabla\times {\bf A}_w).
\end{eqnarray*}

\subsection{Reduction to a one-dimensional model}
\label{sec:one}

Here we assume that the external field ${\bf A}_w$ created by the undulator only depends on the longitudinal variable $z$.
If the beam of electrons has been injected in a proper way, we show below that the motion is exactly described by a one-dimensional Hamiltonian, allowing for a reduced -- but exact -- description of the dynamics. 

This reduction follows from the properties of the Liouville operator $\mathcal{H}=\{ H,\cdot \}$. Recalling that:
\begin{eqnarray*} \frac{\delta H}{\delta f_{\rm mom}}=\Big[1 + {\bf p}^2& +a a^* -i\sqrt{2}(a e^{ikz}{\bf \hat{\mathrm e}}-a^* e^{-ikz}{\bf \hat{\mathrm e}^*}) \cdot {\bf A}_w
\\ &+\vert {\bf A}_w\vert^2-2 {\bf p}_\perp \cdot ({\bf A}_w+{\bf A})\Big]^{1/2},\end{eqnarray*}
the Liouville operator reads:
\begin{eqnarray*} 
\mathcal{H}&=\iint d^3q d^3p &\frac{f_{\rm mom}}{\sqrt{1+{\bf p}^2+({\bf A}+{\bf A}_w)^2}} \Bigg[ \left({\bf p}_\perp-{\bf A}_w-{\bf A}\right) \cdot\frac{\partial}{\partial {\bf q}_\perp} \frac{\delta}{\delta f_{\rm mom}} 
\\ &&+p_z \frac{\partial}{\partial z} \frac{\delta}{\delta f_{\rm mom}} -({\bf A}+{\bf A}_w)\cdot \frac{\partial}{\partial z}({\bf A}+{\bf A}_w) \frac{\partial}{\partial p_z} \frac{\delta}{\delta f_{\rm mom}} \Bigg] 
\\ &&+\frac{i}{kV} \left[ \frac{\partial H}{\partial a} \frac{\partial}{\partial a^*} - \frac{\partial H}{\partial a^*} \frac{\partial}{\partial a} \right]. 
\end{eqnarray*}
The $\frac{\partial}{\partial {\bf p_\perp}} \frac{\delta}{\delta {f_{\rm mom}}}$ term has disappeared since its factor $\partial({\bf A}+{\bf A}_w)/\partial {{\bf q}_\perp}$ vanishes, as the fields are assumed not to depend on the transverse direction ${\bf q_\perp}$. As a consequence, there is no evolution for the particles distribution $f_{\rm mom}({\bf q},{\bf p})$ along the direction ${\bf p_\perp}$. This can be seen by considering a distribution function of the following form
$$
f_{\rm mom}({\bf q},{\bf p})=\hat{f}({\bf q},p_z)\delta({\bf p}_\perp).
$$
Under the Liouville operator, we see that $\mathcal{H}f_{\rm mom}$ is also proportional to $\delta({\bf p}_\perp)$, using an integration by parts and the fact that
$$
\frac{\delta f_{\rm mom}({\bf q},{\bf p})}{\delta f_{\rm mom}({\bf q}',{\bf p}')}=\delta({\bf q}-{\bf q}')\delta({\bf p}-{\bf p}').
$$ 
We recall that ${\bf p}_\perp$ has been translated by ${\bf A}_w$ in Sec.~\ref{sec:can}, so
in other words, if the beam is initially injected with the transverse velocity ${\bf A}_w/\sqrt{1+|{\bf A}_w|^2}$ [from Eq.~(\ref{eq:velocity})], it remains with this specific transverse velocity.
Then, once restricted to the $\delta({\bf p}_\perp)$ distribution, it comes that the transverse profile of $f$ does not act on the longitudinal dynamics any more. In other words, the set $\mathcal{F}$ of observables $F$ which do not depend on the transverse component of the distribution, i.e. $F$ such that
$$\frac{\partial}{\partial {\bf q}_\perp}\frac{\delta F}{\delta \hat{f}}=0, $$
is stable by $\mathcal{H}$ (i.e. $\mathcal{H}F\in \mathcal{F}$ if $F\in \mathcal{F}$), since ${\bf A}$ and ${\bf A}_w$ do not depend on ${\bf q}_\perp$. This allows one to focus on the longitudinal dynamics by defining a reduced dynamics on $\mathcal{F}$, with $\tilde{f}(z,p_z)$ the longitudinal distribution as new variable associated with the particles. The bracket reduces to~:
\begin{eqnarray*} \{ F,G \} &=& \iint dz dp_z \tilde{f} \left[\frac{\partial}{\partial p_z} \frac{\delta F}{\delta \tilde{f}} \frac{\partial}{\partial z} \frac{\delta G}{\delta \tilde{f}} - \frac{\partial}{\partial z} \frac{\delta F}{\delta \tilde{f}} \frac{\partial}{\partial p_z} \frac{\delta G}{\delta \tilde{f}}\right] \\
&&+ \frac{i}{kV} \left(\frac{\partial F}{\partial a} \frac{\partial G}{\partial a^*}-\frac{\partial F}{\partial a^*} \frac{\partial G}{\partial a}  \right), \end{eqnarray*}
and the Hamiltonian becomes
\begin{eqnarray}\label{h1d} H&=& \iint dz dp_z \tilde{f}\sqrt{1 + p_z^2 +a a^* -i\sqrt{2}(a e^{ikz}{\bf \hat{\mathrm e}}-a^* e^{-ikz}{\bf \hat{\mathrm e}^*}) \cdot {\bf A}_w+\vert {\bf A}_w\vert^2}\\
 &&+ k^2 V a a^* -\frac{i kS}{\sqrt{2}} \int dz (a e^{ikz}{\bf \hat{\mathrm e}}-a^* e^{-ikz}{\bf \hat{\mathrm e}^*})\cdot(\nabla\times {\bf A}_w).\nonumber \end{eqnarray}
Since the motion is now one-dimensional, we drop the label $z$ of the momentum in what follows.

\subsection{Particles-field phase frame}
\label{sec:frame}
 
 While Hamiltonian~(\ref{h1d}) is reduced to distribution functions with one dimension (one in space and one in momentum), it still contains some terms which are not specific to the interaction like the term $a a^*$. The emphasis can be put on the interaction between the particles and the wave by considering the dynamics into the particles-field phase frame. We consider a specific medium for the interaction between the particles and the radiated field. For example, we can consider a linear undulator, such as in a Free Electron Laser. The transverse field produced by such an undulator reads~:
\begin{equation} {\bf A}_w=\frac{a_w}{\sqrt{2}} \left( e^{-ik_wz}{\bf \hat{\mathrm e}}+e^{ik_wz}{\bf \hat{\mathrm e}^*} \right).\label{eq:aw} \end{equation}
First we neglect the effects of finite size~: The last term in Eq.~(\ref{h1d}) vanishes since it is a sum of terms proportional to $\int dz e^{\pm i(k+k_w)z}$ terms. The Hamiltonian becomes
$$ H= \iint dz dp \tilde{f}\sqrt{1 + p^2 +a a^* -ia_w(a e^{i(k+k_w)z}-a^* e^{-i((k+k_w)z)})+a_w^2} + k^2 V a a^*. $$
In this Hamiltonian, the last term simply yields the propagation of the electromagnetic wave. Indeed, since $\{ k^2 V aa^*, a \}=-ika$, it generates an $e^{-ikt}$ factor for $a(t)$, and so it is a pure propagation term. This remark calls for a time-dependent change of coordinates. This procedure is fairly standard, e.g., in quantum mechanics, and corresponds to the interaction representation \cite{araki}. 
Since time is not a variable for this model, we first need to extend phase space to add a new pair of canonically conjugate variables with one being similar to time. More precisely, we define the pair of conjugate variables $(\tau,E)$, such that the new Hamiltonian and bracket read
$$ H_{\rm ext}[f,a,a^*,E,\tau]= H[f,a,a^*]+E, $$

\begin{eqnarray*} \{ F,G \} &=& \iint dz dp \tilde{f} \left[\frac{\partial}{\partial p} \frac{\delta F}{\delta \tilde{f}} \frac{\partial}{\partial z} \frac{\delta G}{\delta \tilde{f}} - \frac{\partial}{\partial z} \frac{\delta F}{\delta \tilde{f}} \frac{\partial}{\partial p} \frac{\delta G}{\delta \tilde{f}}\right]\\
&&+ \frac{i}{kV} \left(\frac{\partial F}{\partial a} \frac{\partial G}{\partial a^*}-\frac{\partial F}{\partial a^*} \frac{\partial G}{\partial a}  \right) + \frac{\partial F}{\partial E} \frac{\partial G}{\partial \tau}-\frac{\partial F}{\partial \tau} \frac{\partial G}{\partial E},
\end{eqnarray*}
so that $\dot{\tau}=\{H_{\rm ext},\tau\}=1$, which means that, practically, $\tau$ is identical to the evolution variable $t$.
We now consider the canonical change of variables $(\tilde{f},a,a^*,\tau,E)\rightarrow(\hat{f},\hat{a},\hat{a}^*,\hat{\tau},\hat{E})$ such that 
\begin{eqnarray*}
&& \hat{a}=ae^{ik\tau}, \\
&& \hat{E}=E+ k^2 V aa^*\\
&& \hat{f}(z,p)=\tilde{f}(z,p),\\ 
&& \hat{\tau}=\tau.
\end{eqnarray*}
It can be checked that the symplectic form $ikV da^*\wedge da+dE\wedge d\tau$ is conserved by this transformation. 
This results in a conservation of the Poisson bracket, while the Hamiltonian now reads:
$$ \hat{H}= \iint dz dp \hat{f} \sqrt{1 + p^2 +\hat{a} \hat{a}^* -ia_w(\hat{a} e^{i((k+k_w)z-k\tau)}-\hat{a} e^{-i((k+k_w)z-k\tau)})+a_w^2}+\hat{E}. $$
Finally, the dynamics can be studied in the particles-field phase frame, the latter phase being defined as $\theta=(k+k_w)z-k\tau$, by considering the change of variables 
\begin{eqnarray*}
&& \bar{f}(\theta,p)=\hat{f}(z,p)/(k+k_w),\\ 
&& \bar{a}=\hat{a},\\
&& \bar{E}=\hat{E}+\frac{k}{k+k_w}\iint d\theta dp \bar{f} p,\\
&& \bar{\tau}=\tau, 
\end{eqnarray*}
where $\hat{f}$ has been divided by $k+k_w$ for normalization purposes.
We notice that the term $-k/(k+k_w)\iint d\theta dp \bar{f} p$ has been added to $\hat{E}$ in order to ensure the canonicity of the change of coordinates which translates here into the condition $\{\bar{E},\bar{f}\}=0$. Here we have used the following properties of the functional derivative: If we perform a change of variables of $f$, denoted $\tilde{f}=f\circ g$, then the functional derivative of $\tilde{F}[\tilde{f}]=F[f]$ satisfies
$$
\frac{\delta \tilde{F}}{\delta \tilde{f}}=\frac{\delta F}{\delta f} \left| g'\circ g^{-1}\right|,
$$ 
which is obtained in a straightforward way from the definition of the functional derivative.

Now that we have performed the time-dependent change of coordinates and that the Hamiltonian is time-independent, the $(\bar{\tau},\bar{E})$ variables are somehow artificial and decoupled from the other ones. It is more convenient to work in the reduced space and drop this additional pair of variables. For notation purposes, we drop the bars over the other variables.
The Hamiltonian now reads:
\begin{equation}\label{h77} \bar{H}= \iint d\theta dp \bar{f} \left[\sqrt{1 + p^2+a_w^2 -ia_w(\bar{a} e^{i\theta}-\bar{a}^* e^{-i\theta}) +\bar{a} \bar{a}^*}-\frac{k}{k+k_w}p\right]. \end{equation}
The resulting Poisson bracket writes
\begin{eqnarray} \{ F,G \} &=& (k+k_w)\iint d\theta dp f \left[\frac{\partial}{\partial p} \frac{\delta F}{\delta f} \frac{\partial}{\partial \theta} \frac{\delta G}{\delta f} - \frac{\partial}{\partial \theta} \frac{\delta F}{\delta f} \frac{\partial}{\partial p} \frac{\delta G}{\delta f}\right]\nonumber \\
&&+ \frac{i}{kV} \left(\frac{\partial F}{\partial a} \frac{\partial G}{\partial a^*}-\frac{\partial F}{\partial a^*} \frac{\partial G}{\partial a}  \right)\label{bratheta}. \end{eqnarray}

\subsection{Resonance condition and high-gain amplification}
\label{sec:reso}

The next step is to expand Hamiltonian~(\ref{h77}) around a resonant value.
Following Eq.~(\ref{evol}), Hamiltonian (\ref{h77}) and bracket (\ref{bratheta}) yield the following equations of motion:
\begin{eqnarray*} \frac{df}{dt}&=&-(k+k_w)\left( \frac{p}{\sqrt{1 + p^2+a_w^2 -ia_w(a e^{i\theta}-a^* e^{-i\theta}) +a a^*}}-\frac{k}{k+k_w} \right)\frac{\partial f}{\partial \theta} 
\\ &&+\frac{a_w(k+k_w)(a e^{i\theta}+a^* e^{-i\theta})}{\sqrt{1 + p^2+a_w^2 -ia_w(a e^{i\theta}-a^* e^{-i\theta}) +a a^*}} \frac{\partial f}{\partial p}
\\ \frac{da}{dt}&=&-\frac{k}{V}\iint d\theta dp f\frac{ia-a_w e^{-i\theta}}{\sqrt{1 + p^2+a_w^2 -ia_w(a e^{i\theta}-a^* e^{-i\theta}) +a a^*}}
\end{eqnarray*}
From these equations it can be seen that the system is at equilibrium for $a=0$ and $f(\theta,p)=\delta(p-p_R)F(\theta)$, where $F(\theta)$ is a distribution which satisfies $\int d\theta e^{-i\theta}F(\theta)=0$, and $p_R$ is given by $$\frac{p_R}{\sqrt{1+a_w^2+p_R^2}}-\frac{k}{k+k_w}=0.$$
 This {\it resonant momentum} $p_R$ can be linked to a {\it resonant energy} $\gamma_R$ for the particles, defined by
\begin{equation}\label{gammar} \gamma_R=\sqrt{1+a_w^2+p_R^2}=\sqrt{1+a_w^2}\frac{k+k_w}{\sqrt{k_w(2k+k_w)}}. \end{equation}
In the limit $k \gg k_w$, the definition (\ref{gammar}) of the resonant energy yields the usual definition $\gamma_R=\sqrt{k(1+a_w^2)/(2k_w)}$ (see for example \cite{bonham}).

However, this equilibrium is unstable, and exposed to small perturbations, the wave starts growing and destabilizes the particles at $p=p_R$. This instability is responsible for the high-gain growth of the wave, which is taken advantage of in devices such as FEL.
The dynamics can be linearized around this equilibrium point: Assuming the momenta of the particles remain close from the resonant one $p_R$, we shift $p$ by $p_R$, by defining $\hat{f}(\theta,\hat{p})=f(\theta,p)$ with $\hat{p}=p-p_R$. We also consider that the amplitude of the radiated field is weak compared to the resonant energy
$$ |a|\ll \gamma_R. $$

Then, Hamiltonian (\ref{h77}) expands, at the first order in $a$ and second order in $p$ (we have dropped the hat over $\hat{p}$ and $\hat{f}$), as follows:
$$ H_{\rm lin}=\iint d\theta dp f\left[ \frac{1+a_w^2}{\gamma_R^3}\frac{p^2}{2} -\frac{ia_w}{2\gamma_R} \left( a e^{i\theta}-a^* e^{-i\theta} \right)\right], $$
associated with the bracket~(\ref{bratheta}).

The equations of motion can be normalized through the following change of variables
\begin{eqnarray*}
&& f'(\theta',p')= \frac{1}{\beta}f(\theta=\theta',p=p'/\beta),
\\ && a'= \epsilon a.
\end{eqnarray*}
Moreover, we include a rescaling of time $t'=\alpha t$, i.e. $H'=H/\alpha$, and we consider a new Hamiltonian  $\nu H'$ with a new Poisson bracket ${\nu}^{-1} \{.,.\}$ (which does not change the dynamics). 
Using 
\begin{eqnarray*}
&&\alpha=\frac{1}{\gamma_R}\left(\frac{a_w^2 k_w(k+k_w/2)}{2kV}\right)^{1/3}, \\
&& \beta=\frac{2}{k+k_w} \left(\frac{2kVk_w^2(k+k_w/2)^2}{a_w^2}\right)^{1/3}, \\
&& \epsilon=\left(\frac{4k^2V^2k_w(k+k_w/2)}{a_w}\right)^{1/3}, \\
&& \nu=2\left(\frac{2kVk_w^2(k+k_w/2)}{a_w^2}\right)^{1/3},
\end{eqnarray*} 
the Hamiltonian and the bracket become~: 
\begin{eqnarray}\label{hnorm} H&=& \iint d\theta dp f \frac{p^2}{2}-i \iint d\theta dp f \left(ae^{i\theta}-a^*e^{-i\theta}\right),
\\ \{ F,G \} &=& \iint d\theta dp f \left[\frac{\partial}{\partial p} \frac{\delta F}{\delta f} \frac{\partial}{\partial \theta} \frac{\delta G}{\delta f} - \frac{\partial}{\partial \theta} \frac{\delta F}{\delta f} \frac{\partial}{\partial p} \frac{\delta G}{\delta f}\right] + i \left(\frac{\partial F}{\partial a} \frac{\partial G}{\partial a^*}-\frac{\partial F}{\partial a^*} \frac{\partial G}{\partial a}  \right). \nonumber 
\end{eqnarray}

Finally, the canonical change of variables $(a,a^*)\rightarrow (\phi,I)$ such that $a=\sqrt{I}e^{-i\phi}$ (so that $\{ \phi,I \}=1$) allows one to retrieve Hamiltonian (\ref{eqn:boni}) associated with the bracket~(\ref{eqn:braboni}).

\subsection{Conserved quantities}
\label{sec:cons}

Apart from $H$ as given by Eq.~(\ref{ham0}), the Hamiltonian system of charged particles interacting self-consistently with electromagnetic fields has its total momentum as conserved quantity, as it was reported in Ref.~\cite{bialynicki}:
$$ {\bf P}[f,{\bf E},{\bf B}]=\iint d^3qd^3p f {\bf p} +\int d^3q {\bf E}\times{\bf B}. $$
We perform the same approximations and reductions as the ones done on Hamiltonian (\ref{ham0}). In this way, we recover the conserved quantity of the reduced model. For example, in the canonical formulation (\ref{ham1}) and (\ref{brac1}), the total momentum turns into:
$$ {\bf P}[f_{\rm mom},{\bf A},{\bf Y}]=\iint d^3qd^3p f_{\rm mom} ({\bf p}-{\bf A}) -\int d^3q {\bf Y}\times \left( \nabla\times {\bf A} \right).$$

Furthermore, when considering a monochromatic, circularly polarised plane-wave such as the one given by Eq.~(\ref{approxwave}), the total momentum is decomposed into a transverse and a longitudinal component:
\begin{eqnarray*}
&& {\bf P}_\perp [f,{\bf A}]=\iint d^3qd^3p f ({\bf p_\perp}-{\bf A}-{\bf A}_w),\\
&& P_z [f,{\bf A},{\bf Y}]=\iint d^3qd^3p f p_z +k^2 V a a^* +k\int d^3q \left({\bf A}_w \times {\bf Y}\right)\cdot {\bf \hat{e}}_z.
\end{eqnarray*}
When focusing on the longitudinal dynamics (see Sec.~\ref{sec:one}), the conserved quantities ${\bf P}_\perp$ can be dropped since they provide information on the transverse dynamics. Then, when considering the specific external field (\ref{eq:aw}), if we neglect the finite-size effect, the last term of the longitudinal momentum can be dropped, so that we get, for Hamiltonian (\ref{h77}), the following conserved quantity
$$P_z [f,a,a^*]=\iint d^3qd^3p f p +k^2 V a a^*.$$
Using the same normalization as in Sec.~\ref{sec:reso}, $P_z$ becomes a conserved quantity for Hamiltonian (\ref{hnorm}):
$$ P=\iint d\theta dp f p + aa^*. $$
Finally, using the canonical change of variables $(a,a^*)\mapsto (\phi,I)$, it becomes:
$$ P=\iint d\theta dp f p + I,$$
i.e. the average momentum plus the intensity is conserved by Hamiltonian (\ref{eqn:boni}).

\section*{Conclusion}

In this paper, we derived a reduced Hamiltonian for the interaction between a wave and a beam of charged particles driven by an external field, on the sole assumptions of transverse fields and on-axis injection for the particles. A resonance condition -- around an unstable equilibrium point -- has been identified, allowing for a linearization of the dynamics. Finally, under the extra hypotheses of high-energy particles and weak radiated field, a paradigmatic Hamiltonian has been retrieved within a fully Hamiltonian treatment. The main advantage of the present derivation is a fully algebraic framework which is well suited to include additional effects, like for instance, higher order terms in the expansions, or strategies using other harmonics of the radiated field~\cite{mcne06}. We have shown here that this treatment allows one to recover some general features of the Vlasov-Maxwell equations in a very natural way, like a conserved quantity of the flow.

\section*{Acknowledgements}

We acknowledge useful discussions with G.~De~Ninno, E.~Allaria, D.~Fanelli, Y.~Elskens, R.~Pa\v{s}kauskas, and the Nonlinear Dynamics group at CPT. This work is supported by Euratom/CEA (contract EUR 344-88-1 FUA F).

\end{document}